\documentclass[12pt]{article}
\usepackage{graphicx} % Required for inserting images

\usepackage[utf8]{inputenc} % allow utf-8 input
\usepackage[T1]{fontenc}    % use 8-bit T1 fonts
\usepackage[hidelinks]{hyperref}      % hyperlinks
\usepackage{url}            % simple URL typesetting
\usepackage{booktabs}       % professional-quality tables
\usepackage{amsfonts}       % blackboard math symbols
\usepackage{amsmath}
\usepackage{nicefrac}       % compact symbols for 1/2, etc.
\usepackage{microtype}      % microtypography
\usepackage{lipsum}		    % Can be removed after putting your text content
\usepackage{graphicx}
\usepackage{natbib}
\usepackage{doi}
\usepackage{rotating}
\usepackage{makecell}
\usepackage{xcolor}
\usepackage[normalem]{ulem}
\usepackage{soul}
\usepackage{longtable}
\usepackage{array} 
\usepackage{booktabs}
\usepackage{enumitem} %for controlling enumerate
\usepackage{comment}
\usepackage{caption}
\usepackage{adjustbox}
\usepackage{floatrow}

\usepackage[italian, greek, english]{babel}
\usepackage[LGR, T1]{fontenc}

\usepackage{endnotes}
\let\footnote=\endnote

\title{Edgeworth's exact and naturally weighted evolutionary utilitarianism and the happiness of Mr. Pongo}
\author{Alberto Baccini \footnote{Dipartimento di Economia Politica e Statistica, Università degli Studi di Siena, Italy}}
\date{15 October 2025}

\begin{document}

\maketitle

\begin{abstract}
  This article challenges the conventional reading of Francis Ysidro Edgeworth by reconstructing his intellectual project of unifying the moral sciences through mathematics. The contribution he made in the first phase of his writing, culminating in \textit{Mathematical Psychics}, aimed to reconfigure utilitarianism as an exact science, grounding it in psychophysics and evolutionary biology. In order to solve the utilitarian problem of maximizing pleasure for a given set of sentient beings, he modeled individuals as ``quasi-Fechnerian'' functions, which incorporated their capacity for pleasure as determined by their place in the evolutionary order. The problem of maximization is solved by distributing means according to the individuals' capacity for pleasure. His radical anti-egalitarian conclusions did not stem from an abstract principle of justice, but from the necessity to maximize welfare among naturally unequal beings. This logic was applied not only to sentients of different evolutionary orders, such as Mr. Pongo, a famous gorilla, and humans, but also to human races, sexes, and classes. The system, in essence, uses the apparent neutrality of science to naturalize and justify pre-existing social hierarchies.
  
  This analysis reveals that the subsequent surgical removal of his utilitarianism by economists, starting with Schumpeter, while making his tools palatable, eviscerates his overarching philosophical system
  
  \textbf{Keywords:} Francis Ysidro Edgeworth; Mathematical Psychics; Utilitarianism; Anti-egalitarianism; Model transfer.
\end{abstract}

\section{Introduction}

Francis Ysidro Edgeworth occupies a distinctive place in the history of economic thought. He can be considered as a kind of hero for the Whig approach: a thinker whose scattered contributions anticipated later developments in economics, welfare economics, economics of happiness, bargaining theory, probability, and statistics. Yet -- as \citet{keynes1926} famously observed -- Edgeworth never produced a synthesizing treatise that could stabilize his legacy. The result is an influence that is both pervasive and elusive, channelled through a dispersed body of papers, reviews, and occasional essays.\footnote{The most complete bibliograpy is \citet{baccini2003} including 4 books, 198 journal articles, 204
reviews and 139 entries for the \textit{Dictionary of Political Economy} \citep{palgrave}.
}

The reception of Edgeworth's ideas and tools in subsequent economic literature has been sporadic and fragmented. This selective adoption has obscured the broader intellectual milieu from which they emerged. Moreover, these concepts were often incorporated as neutral, glorified technical apparatuses, divorced from the cultural context that shaped them \citep{scarf1962, shubik1959, yaari, creedy1986, coase1988, colander,  eceiza}. A comprehensive understanding of the origins of many current tools of neoclassical economics necessitates reconstructing this lost context. This historical reconstruction is essential not only for a complete understanding but also for a critical appraisal of modern economic tools. It forces us to ask whether these instruments shed their original ideological connotations to become truly neutral, or the underlying ideological framework has been silently imported alongside them. The case of Edgeworth is emblematic in this regard. His unique intellectual trajectory and cultural breadth consistently intertwined philosophical and foundational concerns with technical innovation. For him, the technical dimension was a direct consequence of the philosophical one; tools and ideas were derived from a prior ideological and ethical commitment. 

Systematic reconstructions of Edgeworth's thought have been relatively rare. Early and influential overviews by Peter \citet{newman1987} mapped Edgeworth’s contributions to economics, while John \citet{creedy1986} situated him squarely within the emergence of the neoclassical program, emphasizing the technical innovations of \textit{Mathematical Psychics} and other related writings. Stephen \citet{stigler1978, stigler1986, stigler1987, stigler1999} highlighted Edgeworth’s pioneering work on statistical inference and ``law of error'', foregrounding a methodological sophistication that prefigured twentieth-century statistics. Alberto \citet{baccini2007} reconstructed the common foundation of Edgeworth's ethics and probability, and proposed to consider the dispersed papers on probability as a proper \textit{Treatise on probabilities} \citep{baccini_2009}.

Philip Mirowski (1994) offers perhaps the most ambitious attempt to integrate Edgeworth’s economics, probability, and statistics into a single narrative about the mathematization of the social sciences. Mirowski’s synthesis has become a reference point for subsequent scholarship. \citet{baccini_2011} followed a similar line to Mirowski but presented a synthetic alternative overview of Edgeworth's work, which forms the background for this paper.

This study is dedicated to reconstructing the subtle links between Edgeworth's ethics, the foundations of utilitarian and economical calculus, and the problem of the indeterminateness of contracts and arbitration. These themes were developed almost exclusively in the first wave of his publications, between 1876 and 1882. Evidence from his later works suggests that Edgeworth himself considered this foundational step concluded \citep{edgeworth1887, Edgeworth_rev_price, Edgeowrth_rev_bonar}, seeing no need to revisit it explicitly. However, the conceptual problems opened by that initial discussion created an intellectual trajectory that led him to subsequently engage with the philosophy of probability and, from there, to his pioneering work in statistics.\footnote{See on this \citet{mirowski1994} and \citet{baccini_2011}.}

The paper is organized into three main sections, which follow the chronological order of the three principal texts from this initial phase of Edgeworth's writing. A final section synthesizes the analysis and presents the concluding arguments.

\section{\textit{New and old methods of ethics}} 

Edgeworth's first book \emph{New and Old Methods of Ethics, or `Physical Ethics' and `Methods of Ethics'} (hereinafter NOME) was published in 1877, when he was 32 years-old \citep{edgeworth1877}. NOME is composed of two sections: the first discusses utilitarian ethics in reference to the works of Henry \citet{Sidgwick} and Alfred \citet{Barrat}; the second proposes and solves the problem of ``exact utilitarianism''.

\textbf{The intellectual milieu and the inspiration.} More generally, Edgeworth's project in NOME emerged from an intellectual \textit{milieu} engaged with evolutionary psychology, psycho-physics, and utilitarianism. The central forum for this debate was \textit{Mind}, the pioneering journal founded by Alexander Bain in 1876 that carved out a unique intellectual space at the intersection of philosophy, psychology, and the sciences of the mind \citep{Staley}.

The key inflection point was possibly Edgeworth's collaboration with his friend, the psychologist James Sully \citep{block, ryan},  whom he assisted in writing his book \textit{Pessimism. A History and a Criticism} \citep{sully}.\footnote{ ``I have great pleasure in acknowledging the assistance lent me by my friend Mr. F. Y. Edgeworth, M.A., of Balliol College, to whose careful perusal of the proof-sheets I am indebted for numerous improvements both in the argument and in the style of the work'' \citep[preface page]{sully}.} 
In the two books \citet{sully_1874, sully} had written at that time, he defended the physical basis of evolutionary psychology in reference to the works of German psycho-physics. In \textit{Sensation and Intuition}, \citet{sully_1874} engaged with the Spencerian theory of mental evolution, which posited that mental capacities are inherited through the hereditary transmission of cerebral structures.\footnote{ ``The hypothesis of evolution, besides explaining the general nervous conditions of mental phenomena in the human organism, seeks to account for some of those modes of feeling which, as observed, do not appear to be the effects of causes acting within the limits of the individual life . And this it does, as I have hinted, by help of the theory of hereditary transmission. According to this hypothesis, a man's experiences and habits, while they distinctly modify his own cerebral structure and mental capacity, tend also to modify those of his offspring'' \citep[p. 5]{sully_1874}.} 
At the same time, he plainly presented ``the recent German experiments'' on stimulus and sensation that of Wilhelm \citet{wundt} and Gustav \citet{fechner}, that ``constitute in a peculiar manner the borderland of physiology and psychology'' \citep[p. 36]{sully_1874}.

It was likely in the works of Sully that Edgeworth identified the starting point for his own research program.  Specifically, he was drawn to Sully's account of Herbert Spencer's evolutionary speculation. According to \citet[p. 3]{sully_1874}, Spencer speculated that, regardless of their precise physiological basis, evolution must have forged a vital link: pleasurable feelings are associated with conditions and actions beneficial to the organism, while painful ones are linked to those that are harmful. However, according to Sully: 
\begin{quote}
``Now a complete scientific doctrine of pleasure and pain, in other words, a systematic science of hedonics, has, as yet, no existence.'' \citep[p. 264]{sully}
\end{quote} 

\textbf{A physical foundation for ethics.} These quotations from Sully contain all the fundamental elements of Edgeworth's NOME. The core issue of NOME's first section is how human actions are conceptualized within moral philosophy \citep{baccini2007, Yee}: can such actions be simplified to mere ‘contractions’ (the pursuit of pleasure) and ‘irritations’ (the avoidance of pain) within our nervous system, or do they belong to the realm of consciousness? Edgeworth found Barratt's reductionist stance -- which posits a physical origin for all human conduct in pleasure -- highly compelling and was inclined to agree that actions have a basis in human physiology. Nevertheless, individuals perceive a distinction between reflexive and volitional acts, as well as between choices made for hedonistic reasons and those made for other motives. Consequently, it is challenging to uphold Barratt's position that every action stems directly from pleasure, just as it is difficult to maintain Sidgwick's view that human actions originate solely in the ‘idea’ of what is pleasant.

Edgeworth's solution was to synthesize the perspectives of Barratt and Sidgwick. He proposed that the body and mind are inextricably linked, with every change in the material substrate of the mind corresponding to a mental event. The pursuit of pleasure and avoidance of pain can be explained by the nervous system forming interconnections that link present sensations to past experiences of pleasure and pain, which may be more or less distant in time. These experiences are of two types: those personally acquired by the individual over their lifetime, and those transmitted to the individual by their species through the inherited physical structures of the nervous system.

The acknowledgment of this second type of experience originates in the psychological work of \citet{spencer1870}, as previously seen in Sully's appraisal of him. More specifically, Spencer argued that humans are born with a nervous system containing pre-defined connections passed down from the species. These inherited connections create a ``preparedness to cognize'' in the individual. The nervous system is thus an organic, evolving entity that incorporates a priori knowledge within its very structure. While individual experience provides the concrete material essential for all thought, without it, the nervous system's innate organization would not independently generate knowledge.

For \citet[p.10]{edgeworth1877}, the physical architecture of the human brain -- which stores both the historical experience of the species and the immediate experience of the individual -- forms the basis for asserting that ``definite physical phenomena (which Mr. Barratt (…) calls pleasure) are the cause of all human action''. This constitutes the physical foundation of ethics: every action can be traced back either to an individual's personal experiences of pleasure and pain or to experiences genetically inherited from the species. This framework allows for the justification of seemingly non-hedonistic actions, by linking them to an ancestral experience of pleasure and pain. The crucial concept for this ethical foundation was the expansion of the domain of pleasure experience to include that which is undertaken by the species and genetically passed to subsequent generations. 

Edgeworth aimed to establish a scientific foundation for utilitarianism by grounding moral philosophy in the empirical realities of human physiology. His project sought to identify the physical mechanisms that make ethical behaviour possible, moving the discourse from abstract speculation to observable fact. 

\textbf{Material origin of sympathy.} In this context, Edgeworth called for a physiological investigation into the material origins of sympathy \footnote{\citet{lanzoni} reconstructs the discussion about sympathy in \textit{Mind}.}, which he believed was the key to understanding altruistic action. This is encapsulated in the central passage he later termed the reconciliation between \textit{egoism} and \textit{altruism}'' \citep[p. 104]{edgeworth1881}: \begin{quote}Physiology could discover the cerebral cause of non-hedonistic action which is directed to the pleasure of others (corresponding to sympathy in Mr. Bain's sense)'' \citep[p.14]{edgeworth1877}.
\end{quote}
By locating the cerebral cause of sympathy, Edgeworth's programme promised to transform altruism from a metaphysical puzzle into a scientifically explicable phenomenon. This reconciliation was not merely philosophical; it was designed to provide a robust basis for justifying the adoption, at an individual level, of the aim of exact utilitarianism. Indeed,
\begin{quote}
the utilitarian end is the greatest possible quantity of happiness of sentient beings, without regard to their number or distribution” \citep[p. 35]{edgeworth1877}.
\end{quote} 

At the individual level, the utilitarian end may demand individual ``self-sacrifice'', a direct subtraction from one's own welfare to benefit the welfare of others. Within Edgeworth's programme this demand loses its paradoxical character. It is conceived as the expression of an identifiable psycho-physiological mechanism, through which the pleasure of others becomes a neurologically-grounded and rational motivation for the individual. Edgeworth's solution was not to validate all non-hedonistic actions as rational, but to demonstrate that ``only one of non-hedonistic impulse -- that toward the happiness of others -- is rational'' \citep[p. 16]{edgeworth1877}. By grounding sympathy in observable physical and psychological conditions, Edgeworth argued it could be elevated to the status of the sole non-hedonistic motive that satisfies Sidgwick's stringent requirement for rational action \citep[pp. 15-16]{edgeworth1877}. 

This culminated in a vision of the rational utilitarian agent who acts though a complete identification with the pleasure of all others:
\begin{quote}
``It is possible to conceive a soul moved only by the attraction of hedonism, so moved that the integrated attractive force of the pleasures of all other souls, an attraction not varying with the social distance, should constitute an acceleration in the direction of utilitarianism, compared with which any other motive, any other hedonistic force whatsoever, might be neglected; or rather could only act in directions not interfering with the great resultant.'' \citep[28]{edgeworth1877}
\end{quote}
According to Edgeworth, a desire for universal happiness, though not yet widespread, is nascent among the clever (\foreignlanguage{greek}{χαρίεντες}) and morally serious (\foreignlanguage{greek}{σπουδαῖοι}) who form an avant-garde in the course of evolution. Consequently, an enlightened egoist, recognizing this evolutionary trajectory, is presented with a powerful motive and the capacity to cultivate altruistic desires. The ultimate aim is not to achieve pure utilitarianism -- that Sidgwick called ``universalistic utilitarianism'' -- but to asymptotically approximate it, aligning self-interest with the moral direction of societal evolution \citep[33]{edgeworth1877}. 

\textbf{The intellectual origin of exact utilitarianism.} After having defined universalistic utilitarianism as the foundation of ethics,\footnote{The precise quotation is: ``taking for granted as the standard of morals utilitarianism'' \citep[p. 35]{edgeworth1877}. According to \citet[89]{newman1987}, it is precisely from this point onward that NOME takes off for economists.} Edgeworth moved on to the next problem, that of utilitarian distribution.   

%\textcolor{green}{
%The nature of the pleasure machine examined in NAME is defined by everything outlined above. It is important to note that not all such machines are identical; they vary in their capacity for happiness and their sensitivity thresholds for pleasure. Therefore, the solutions to utilitarian problems of maximizing social welfare depend on the distribution of these sentient beings within a population and their differing characteristics. However, Edgeworth ultimately concluded that ‘[w]ith regard to the theory of distribution, there is no indication that, at any rate between classes so nearly in the same order of evolution as the modern Aryan races, a law of distribution other than equality is to be wished’ (Edgeworth 1877: 78). The rationale for this concluding statement is unclear: how can the plurality of diverse agents be synthetically represented as a single, uniform pleasure machine with identical and ? More broadly, what is the epistemological status of the very concept of a human as a pleasure machine?}

As already anticipated, according to Edgeworth, the ultimate aim of utilitarian ethics is the maximization of collective welfare: the utilitarian end is defined as “the greatest possible quantity of happiness of sentient beings, without regard to their number or distribution” \citep[p. 35]{edgeworth1877}. The central problem, then, is how to allocate, among a given set of sentients, a fixed amount of stimulus -- itself tied to a given quantity of material resources -- so as to generate the maximum aggregate happiness \citep[p. 40]{edgeworth1877}.

Edgeworth defined the problem as ``exact utilitarianism'' and cleary stated his intellectual debt toward Sidgwick and, rather surprisingly, toward Fechner: 
\begin{quote}
``Not differently, though not quite so distinctly, is the utilitarian end defined by Fechner in his charming treatise \textit{Ueber das hochste Gut} [\textit{On the Greatest good}]. The problem of distribution Fechner illustrates by the problem to divide a given number into a given number of parts, so that the product of the parts should be a maximum.'' \citep[pp. 35-36]{edgeworth1877}   
\end{quote}
In the preceding quotation Edgeworth referred to a Fechner's booklet \citep{fechner} in which ``he sketched a naturalistic ethics and
tried to draw some ethical conclusions from his doctrine of immortality'' \citep[p. 51]{heidelberger}. 

The passage from Fechner's text referenced by Edgeworth concerns the principle that the pursuit of maximum collective pleasure necessarily requires a specific optimal distribution among individuals. Fechner illutrated this idea with a mathematical analogy: just as the maximum product of the division of a number in a given number of parts (e.g., 12 divided in two parts) is achieved through equal parts ($6\times6=36$), the total pleasure is maximized when distributed equally. However, he specified that in the social realm, this equity must be adapted to individual differences such as merit, education, and status, thereby justifying an unequal distribution when it serves the maximum general welfare, thus uniting ``expediency'' and justice. As we will see, Edgeworth's solution to the distributive problem will be very akin to Fechner's argument.\footnote{The influential line connecting this text, in which Fechner had anticipated results later formalized by Edgeworth in NOME, has never been sufficiently emphasized by previous commentators. For this reason, we provide the full quotation from the German text, automatically translated: 
\begin{quote}
``Since our maximum principle demands the greatest possible pleasure among people, it also automatically demands its distribution among people. One could say: pleasure increases, and pain diminishes, up to a certain point, with its distribution. The poet expressed it succinctly and beautifully with these words: shared joy is double joy, shared pain is half pain. 
Not every distribution is indifferent, however, and our principle also determines the correct distribution. Generally speaking, one can see the correct distribution itself as a source of pleasure, and the incorrect one as a source of displeasure. Thus, insofar as our principle demands the maximum of pleasure, it simultaneously determines as the most correct distribution precisely that which is the condition of this maximum.
I will explain this with a mathematical analogy: the product of the parts into which a number can be divided depends on the way it is divided, a function of how one expresses oneself. The largest possible product always belongs to a single, specific way of division. For example, if I divide 12 into 1 and 11, the product of the two parts is 11; if I divide it into 2 and 10, it is larger, namely 20; if I divide it into 3 and 9, it is again larger, namely 27. etc. The most advantageous division is into equal parts, namely 6 and 6, which gives the maximum product of 36. If it were necessary to divide not into two but into three parts, then the three equal parts 4, 4, 4 would also give the maximum product, namely 64. And so in general, for any size and number of parts that one may choose, the equal division is the most advantageous for obtaining the highest product, while the unequal division is all the more advantageous the closer it approaches the equal division.
Similarly, the magnitude of pleasure in humanity as a whole is a function of the way it is distributed among its individual members, namely in such a way that, if everything in these members were equalized, the maximum pleasure would be expected from an equal distribution among them. However, not everything is equal among them, and this is linked to the expediency and justice of distribution in particular. Disposition, character, education, merit, and inherited or acquired status of people determine that it is better to accumulate, or leave accumulated, more pleasure or pleasure-giving substances on one than on another'' \citep{fechner_good}.  
\end{quote}
}

\textbf{Maximising social welfare.} The physical foundations of ethics allowed Edgeworth to represent a sentient through a ``quasi-Fechnerian'' function and use advanced mathematics to solve the problem of maximization of social welfare.\footnote{\citet{newman1987} provided a synthesis of NOME in modern notation; but see also \citet{wall} and \citet{Yee}.} It is worth noting the universalistic approach with which the problem is defined: the sentient beings among whom the stimulus must be distributed are not defined as a static population of human beings, but, following Sidgwick, as a multigenerational population of living beings, including human beings.

To solve the problem, Edgeworth proposed to represent each sentient through an increasing concave quasi-Fechnerian function. The amount of pleasure they experience increases with the level of stimulus, but the rate of this increase slows down as the stimulus gets higher. More precisely, Edgeworth adopted a quasi-Fechnerian function 
\begin{equation}
 \pi_s = k_s ( f(y_s) - f(\beta_s)),   
\end{equation} where $\pi_s$ indicates the pleasure of the sentient $s$, $y$ the quantity of the stimulus they receive, $\beta_s$ their ``threshold, the lowest values of stimulus for which there is sense of pleasure at all'' \citep[p. 40]{edgeworth1877} and $k_s$ is a variable denoting the individual ``capacity of pleasure''. This latter notion is defined as \begin{quote}``A sensory element is said to have greater capacity for pleasure, when it not only affords a greater quantity of pleasure, pleasure arising from simple sensations, for the same quantity of stimulus, ceteris paribus; but also a greater increment of pleasure for the same increment of stimulus, ceteris paribus''. \citep[p. 11]{edgeworth1877}  
\end{quote}
It is worthwhile to highlight that this quasi-Fechnerian function is merely a descriptive representation or a model of sentient beings, and not a utility function to be maximized by rational agents.\footnote{As pointed out by  \citet[p. 17]{mirowski1994}: ``Only by ignoring all the relevant parameters and concentrating singlemindedly upon the function $f$ can a modern neoclassical economist successfully discern her beloved `utility function' in all this''.} 

The ``pleasure of the whole'' is therefore computed as the weighted sum of individual pleasures, where the weights are the capacities for pleasure: 
\begin{equation}
 \sum_{s=1}^{S} k_s (f(y_s) - f(\beta_s)).   
\end{equation}
Crucially, Edgeworth argues that the capacity for pleasure ($k_s$) of different sentient beings depends fundamentally on their position in what he calls the ``order of evolution'' \citep[pp. 54-55]{edgeworth1877}. This evolutionary hierarchy assigns differential weights in the calculation of social welfare based on the supposed developmental advancement of different species and different human groups. Indeed, according to \citet[55]{edgeworth1877}, ``sound utilitarianism'' never intended that a gorilla as the then famous Mr. Pongo  ``was to count for one'' in total welfare.

The idea of equality, as defended by John Stuart \citet[93]{Mill} is only a presumption due to ``taking for granted that there is no \textit{material }difference between human creatures'' \citep[55]{edgeworth1877}. Edgeworth  argued that if utilitarians were to accept the Aristotelian premise of a natural hierarchy among humans, such as the existence of the ``natural slaves'' (\foreignlanguage{greek}{φύσει δοῦλος}) then their own principle of maximizing aggregate utility would logically compel them to endorse a privileged class \citep[55]{edgeworth1877}. If such a hierarchy exists a weighted utilitarian welfare is completely justified.

In this passage, Edgeworth's reasoning was rather cautious by conceding that differences among ``human creatures'' existed more conceivably in past than in the present ``stage of the world's evolution'' \citep[p. 55]{edgeworth1877}. Twenty pages later he was more assertive in stating that the cultivated man, ``in virtue of his better education, of  \foreignlanguage{greek}{μουσικὴ} [music] and \foreignlanguage{greek}{γυμναστική} [phisical activities], would possess an advantage, not so much over the savage, as over the pauper'' \citep[p. 74]{edgeworth1877}.

\textbf{An architecture for inequality.} It can be supposed that Edgeworth's universalistic, psychophysical approach was designed to build a social welfare function that embedded inequality into its very architecture. He introduced the concept of natural weights to compare the pleasure of different species of sentients, a move that appeared more objective and less contentious. This move was then functional to justifying the application of similar differential weights among different kinds of humans, based on the same ``order of evolution''. This methodological step made the ensuing unequal consideration of human pleasures within the utilitarian calculus appear as a natural and objective extension of a broader scientific evolutionary principle. By quantifying the capacity for pleasure across various species and among humans themselves, his model explicitly sanctioned their unequal consideration within the utilitarian calculus from the outset. This foundational move, which directly countered John Stuart Mill's call for egalitarian weights, established a justification for disparate treatment prior to any decision on resource distribution. This architecture, privileging certain beings as inherently more capable of happiness and therefore more deserving, contains the conceptual seeds of eugenic thinking. 

According to \citet[p. 91]{newman1987}, Edgeworth ``trasmuted'' his ``happy utilitarian world'', characterized by ``merely [...] individual differences in capacity of pleasure'', into the ``deadly earnestness of Herbert Spencer and Social Darwinism, of Galton and eugenics''. This interpretation suggests that Edgeworth superimposed an ideological framework onto a neutral utilitarian structure. In contrast, the interpretation proposed here contends that the very structure of his reasoning was poisoned by a eugenicist agenda.

\textbf{Maximization of pleasure.} On the basis of the social welfare function, Edgeworth explored the problem of maximization of pleasure. It consider a given set of sentients and a given quantity of stimulus:
\begin{quote}
    ``Given a certain quantity of stimulus to be distributed among a given set of sentients (with the condition that every element is to have \textit{some} stimulus), to find the law of distribution productive of the greatest quantity of pleasure.'' \citep[p. 43]{edgeworth1877}
\end{quote}
He solved the maximization problem in four cases: two ``where the sentients do not differ in the order of evolution'', and two where they differ. More precisely, in the first case sentients have the same capacity of pleasure ($k_s=k$ for all $s$) and the same sensibility ($\beta_s=\beta$ for all $s$); in the second they have the same capacity but different sensibilities.
The solution of the problems was reached by using an ``untrodden method'' \citep[p. 77]{edgeworth1877}, that is a new and appropriate mathematical machinery, reconstructed and celebrated by \citet{creedy1986} and \citet{newman1987}. 

Edgeworth initially developed the solutions to the problems based on ``simple sensations'' and generic quantity of stimulus. He then reformulated the solutions by using the more realistic notion of quantity of ``means to stimulus'' \citep[p. 43]{edgeworth1877} such as ``wealth, power, $\&$c.'', and by explicitly incorporating the order of evolution. According to this second formulation, the optimal solution for the two cases in which sentients do not differ in their order of evolution ``is an equal division of material means'' \citep[p. iii]{edgeworth1877}. Edgeworth highlighted that these two egalitarian outcomes -- where every sentient effectively counts for one -- are merely contingent results of the utilitarian maximization problem. Crucially, and in stark contrast to Mill, the egalitarian doctrine ``is \textit{not} involved in the very meaning of utility'' \citep[p. 54]{edgeworth1877}, it is simply the specific result of the maximization problem under the condition that sentients do not differ in their order of evolution, i.e. when their sensitivity threshold ($\beta$) and capacity of pleasure ($k$) are identical for all.
Conversely, in the two cases where sentients differ in their order of evolution ``the best outcome involves an unequal allocation, assigning more resources to those with the greatest `felicific power''' \citep[p. iii] {edgeworth1877}.

%In the second part of the second section of NOME, Edgeworth generalized the stimulus function  introduced a generalization qui ci andrebbe messa la generalizzazione di p. 57 riportata anche in newman 1987 p. 91

Elaborating on his previous results in the second part of NOME's second section, Edgeworth enlarged the analysis to address the reproduction over time of the utilitarian society and the problem of population. By doing this, he introduced arguments that served to qualify and moderate the strong anti-egalitarianism he had initially defended. In the concluding section of NOME, Edgeworth synthesized the implications of his ``exact utilitarianism'' for distribution and population theory. He argued that while equality is optimal for groups at a similar evolutionary stage, privilege is justified for ``more highly evolved'' classes when a significant evolutionary gap exists: 
\begin{quote}
    ``With regard to the theory of distribution, there is no indication that, at any rate between classes so nearly in the same order of evolution as the modern Aryan races, a law of distribution other than equality is to be wished. The more highly evolved class is to be privileged when there is a great interval, as there is between man and ape, as there may have been between the ranks and races of the ancient world.'' \citep[p. 78]{edgeworth1877}
\end{quote}
On population, he advocated for maximizing the quality -- that is the evolutionary level -- of sentients, but not at the ultimate expense of their quantity. Edgeworth introduced the idea that prior to his main Fechnerian analysis of pleasure there is a critical material constraint: the ``threshold of the necessities of life''. According to his famous quote: 
\begin{quote}
``a hundred philosophers would elicit more happiness than a hundred capuchin monkeys. But perhaps the material which would keep a hundred little monkeys in health and happiness would not feed twenty philosophers!'' \citep[p. 76]{edgeworth1877}
\end{quote} 
Edgeworth argued that before any comparison of capacity for pleasure between beings can be made, a prior condition must be met: each sentient must have enough resources for basic survival. A being can only experience higher pleasures once this biological threshold is crossed. This concept materially limits the application of his utilitarian calculus, as distributing resources to maximize pleasure first requires ensuring that the recipients can actually survive on them.
After this tortuous reasoning, Edgeworth rejected both the prospect that the earth should be inherited either by a small élite (``the most cultivated coterie'') or by a ``large proletariat''. Instead, he envisioned that a ``happy middle class'' -- an optimal blend of evolutionary quality and numerical quantity -- would inherit the earth. 

This conclusion confirms how Edgeworth’s utilitarian framework is fundamentally structured by a hierarchical evolutionist ideology that naturalizes social inequality. It was only after a long theoretical journey that he was able to temper the radical inegalitarian outcomes of his analytical framework, reconciling it with conclusions that ``are acceptable to common utilitarianism, and, if not to common, at least to good, sense'' \citep[p.78]{edgeworth1877}. 

\section{The hedonical calculus}

Two years later, in 1879, Edgeworth published in \textit{Mind} an article entitled ``The hedonical calculus'' \citep{edgeworth1879}, which would later be incorporated without modifications into the second part of \textit{Mathematical Psychics}.

According to \citet[p. 92]{newman1987} it is ``unfortunate'' that it has been the primary source for illustrating Edgeworth's utilitarianism, as he finds it ``both less subtle and more tarnished with eugenics'' than NOME.  \citet[p. 19]{mirowski1994} considered it as ``another assault on the citadel [of the will]'' following the failure of NOME to find a wide audience.\footnote{The only review of the book appeared on \textit{Mind} probably signed by James Sully \citep{review_nome}.} 

The problem tackled in this paper is again the problem of maximization of social welfare that is developed now in relation not only to the distribution of given means of pleasure, but also to the distribution of labour and to the conditions of its reproduction.  

Edgeworth reiterated the definition of individual ``capacity for happiness'' and paired it with the notion of individual ``capacity for work''. This latter is defined as follows:
\begin{quote}
``An individual has more capacity for work than another, when for the same amount whatsoever of work done he incurs a less amount of fatigue, and also for the same increment (to the same amount) whatsoever of work done a less increment of fatigue.'' \citep[p. 395-396]{edgeworth1879}
\end{quote}
He then introduced a postulate stating that 
\begin{quote}
``capacity for pleasure and capacity for work generally speaking go together; that they both rise with evolution'' \citep[p. 399]{edgeworth1879},
\end{quote}
adding that ``even at present we can roughly discriminate capacity for happiness''. 

Edgeworth considered two main objections: (i) capacity is not precisely measurable and (ii) it is not natural but ``artificial, being due to education''. In response to the first objection, Edgeworth conceded that current methods allowed only for a rough discrimination of capacity, but appealed to the authority of Sully 
and \citet{Barrat}, and to the promise of ``hedonimetry'', to bridge this gap. To the second and more damaging objection he simply dismissed it ``in face of what is known about heredity'' \citep[p. 395]{edgeworth1879}. As in NOME, he naturalized differences of capacities as due to heredity. 

Unlike in NOME, he now explicitly introduced a segmentation of individuals in ``sections'' composed of those with a same capacity of pleasure or work. These sections are ranked according to the capacity of pleasure or work of the individuals constituting them. When the distributive problems of means and labour were treated separately, Edgeworth reiterated the solution reached in NOME:  
\begin{quote}
    ``the distribution of means as between the equally capable of pleasure [work] is equality; and generally is such that the more capable of pleasure [work] shall have more means [do more work].'' \citep[p. 398-399]{edgeworth1879} 
\end{quote}
Hence, inside each section, the distribution of means or labour is equality, while between sections it is inequality, with sections with more capable individuals receiving more means or doing more work. 
He then generalized the solution to the case in which means and labour are allocated simultaneously, employing the same mathematical machinery used in NOME. The result confirmed that the maximization of total welfare is achieved through an unequal distribution between sections of society, with higher-ranked sections receiving both more means and a greater allocation of labour.

\textbf{Education policy.} The first direct application of this more complex inegalitarian framework was to education policy. As we have seen, Edgeworth denied that individual capacities of pleasure and work are the result of education. This move is essential for addressing the problem of distributing the ``means of education''. 
Since there is a fundamental correlation between capacity for pleasure, capacity for work, and evolutionary level, he concluded that while raising the evolutionary level of the entire population is desirable, the optimal allocation of a given amount of means of education is not an equal one. Instead, he argued that investment must be concentrated on the ``most advanced'' classes, as they were deemed ``most capable of education and improvement''. This efficiency-based argument led to the conclusion that ``in the general advance the most advanced should advance most''. \citep[p. 400]{edgeworth1879} 

This passage reveals the core of Edgeworth's ideological project: to provide a utilitarian justification for entrenched social hierarchies. He identifies birth, talent, and the male sex as the foundations of ``aristocratical privilege'', a sentiment he traced to the ``wisest of the ancients''. Crucially, Edgeworth argued this sentiment has a ``ground of utilitarianism in supposed differences of capacity''. He seamlessly merged this elitism with his evolutionary framework, asserting that ``capacity for pleasure is a property of evolution'', thereby framing civilization itself as a hierarchical order. This leaded him to a direct defense of historical class structure: the upper ranks, distinguished by their ``grace of life'' and ``courage'', ``not unreasonably received the means to enjoy and to transmit'' their privilege. Conversely, he posited a natural division of labor where lower classes are assigned the work they seem ``most capable'' of, while the work of the higher classes is deemed qualitatively different and thus beyond comparison in severity. Ultimately, Edgeworth moved beyond a utilitarian apology for inequality as a regrettable necessity; instead, he championed it as a positive good, rooted in a natural order of differential value and evolutionary merit (``\foreignlanguage{greek}{axia}'').

\textbf{Political constraint for unsustainable utilitarism.} As in NOME, in this paper Edgeworth also attempted to avoid the most radical consequences of his own conceptual apparatus. He directly confronted the difficulty by asking: “What is the fortune of the least favoured class in the Utilitarian community?” \citep[p. 402, italics in original]{edgeworth1879}. The response was ``\textit{the condition of the least favoured class is positive happiness}'' \citep[p. 403, italics in original]{edgeworth1879}.

This critical question, however, must not be interpreted in a Rawlsian fashion, as suggested by \citet{newman1987}. For Edgeworth, the ``least favoured class'' does not constitute a social group with claims grounded in the rights of a democratic society. Like Bentham, Edgeworth was profoundly dismissive of individual rights, which he derided as demagogical\footnote{``The demagogue, of course, will make short work of the matter, laying down some metaphysical 'rights of man''' \citep[p.129]{edgeworth1881}.} or as a "metaphysical incubus" \citep{edg_rev_jevons}.

Instead, his concern refers to individuals deemed to possess a naturally inferior capacity for pleasure. Consequently, the impetus behind Edgeworth's question is not a commitment to fairness among free and equal citizens, but stems from a calculus of social efficiency and stability. He explicitly argues that while utility maximization could theoretically justify significant sacrifice from the lower classes, there exists a pragmatic lower bound:

\begin{quote}
``In fact the happiness of some of the lower classes may be sacrificed to that of the higher classes. However [...] the required limit is above the starving-point; both because in the neighbourhood of that point there would be no work done, and -- before that consideration should come into force and above it -- because the pleasures of the most favoured could not weigh much against the privations of the least favoured.'' \citep[p. 404]{edgeworth1879}
\end{quote}

This logic of efficiency is complemented by a deeper concern for social preservation. Edgeworth acknowledges that a ``zero of happiness'' is politically unsustainable, necessitating an extra-utilitarian limit:

\begin{quote}
``It may be admitted however that a limit below the zero of happiness, even if abstractedly desirable, would not be humanly attainable; whether because discomfort in the lower classes produces political instability [...], or because only through the comfort of the lower classes can population be checked from sinking to the starving-point [...]. Let politics and political economy fix some such limit above zero.'' \citep[p. 404]{edgeworth1879}
\end{quote}

Thus, Edgeworth's effort to soften the harsh conclusions of his own hedonical calculus required introducing a political constraint from outside his theoretical system. This limitation was born not from a recognition of individual rights, but from a pragmatic concern for economic functionality and social stability and social reproduction within the utilitarian framework.

\citet[pp. 92-93]{newman1987} interpreted ``The Hedonical calculus'' as if Edgeworth abandoned the more egalitarian stance of NOME for a ``poisonous brew'' of utilitarianism and eugenics that would that would then be carried over into \textit{Mathematical Psychics}. This is a generous interpretation. In reality, the very intellectual structure of NOME is poisoned from its foundational axiom of a variable capacity for pleasure. The consequences of this foundational axiom were only temporarily masked in NOME by the long theoretical journey that we have previously reconstructed. The clear eugenic conclusion of ``The Hedonical calculus'' is, therefore, not an aberration; it is the premise revealed.

Consequently, the ``hard and fast line'' of social protection in ``The Hedonical calculus'' should not be misread as a genuine concession to egalitarianism or welfare. It is, rather, the pragmatic containment of a logic of inherent inequality that was present from the very beginning. This limit is not merely paternalistic but structural and functional: it represents the minimum physiological threshold required for the reproduction of the labour force and the stability of the societal (capitalist) order itself. Edgeworth’s system thus acknowledges a brutal, extra-utilitarian necessity: the recognition that even a system premised on natural hierarchy must ensure the bare survival of its least favoured members to guarantee its own perpetuation.

\section{\textit{Mathematical psychics}}
In the two years after the publication of ``The Hedonical calculus'', a significant shift occurred in Edgeworth's intellectual trajectory, a process that \citet[p. 19]{mirowski1994} has termed ``the transmogrification of Edgeworth from a moralist into an economist''. This transition from utilitarian ethics to economics is widely attributed to the influence of William Stanley Jevons (for all \citet{barbe}), who eventually became Edgeworth's neighbour in Hampstead, London. Their association was both intellectual and personal: \citet[p. 181]{sully_my_life} who also moved to Hampstead in 1878, recalled that
\begin{quote}
    ``a common interest drew him [Jevons] and my chum [Edgeworth] together, and so we made a trio in many a pleasant walk and skating excursion''.
\end{quote}

In reality, the scope of Edgeworth's project in \textit{Mathematical Psychics} \citep{edgeworth1881} was far broader than economics alone, as indicated by its revealing subtitle: ``An Essay on the Application of Mathematics to the Moral Sciences''. This term referred to a well-defined group of subjects in the Victorian academy -- most notably at Cambridge, where the Moral Sciences Tripos encompassed psychology, logic and methodology, ethics and political economy. Edgeworth's ambition was thus to unify at least ethics and economics under a single, mathematical framework. 

The new problem for \textit{Mathematical Psychics} was to determine how human beings -- not the more general set of sentient beings --, now modelled as pleasure-maximizing agents capable of decision-making, would interact with one another. The central theoretical challenge was to verify whether the distributive outcome emerging from the interactions of these pleasure machines would be consistent with the utilitarian optimum previously calculated by treating them as passive receptacles of pleasure through the quasi-Fechnerian law of NOME. In essence, Edgeworth sought to prove that a universe of freely contracting, hedonistic individuals would naturally tend toward the very same maximized distributive arrangement dictated by his exact utilitarianism.

The powerful analytical tools that Edgeworth developed in \textit{Mathematical Psychics} -- most notably the generalized utility function, indifference curves, and the diagram that will give rise to Edgeworth's box -- proved so compelling that they came to define his legacy, obscuring the book's broader utilitarian framework. While his ambition was to establish a mathematical foundation for the moral sciences, the influence and enduring utility of these microeconomic constructs shifted the focus of commentators\footnote{With the notable exception of \citet{newman1987} and \citet{mirowski1994}.} onto the mechanics of exchange, a shift that began with Alfred Marshall's famous review \citep{Marshall}. In it, Marshall was the first to dissect and dismiss the book's overarching utilitarian design, thereby obscuring its broader philosophical framework. 

\textbf{Economical and utilitarian calculus.} \citet[pp. 15-16]{edgeworth1881} established a fundamental distinction within the``calculus of pleasure'', separating it into ``Economics'' and ``Utilitarian ethics''.
This theoretical subdivision is directly mirrored in the structure of the second part of \textit{Mathematical Psychics}. Significantly, the second division of this part reproduces the text of his earlier essay,``The Hedonical Calculus'', now re-titled ``Utilitarian Calculus'' with only minor additions in the form of a few footnotes.

According to \citet[p. 15]{edgeworth1881}, ``Economical calculus investigates the equilibrium of a system of hedonic forces each tending to maximum individual utility'', that is a system where each agent strives to maximize their own individual utility. The ``Utilitarian Calculus'', in contrast, investigates ``the equilibrium of a system in which each and all tend to maximum universal utility'', that is a system where each and all agents tend toward maximum universal utility. \citet[pp. 16]{edgeworth1881} contrasted the two species of agents as corresponding respectively to Sidgwick's Egoistic and Universalistic Hedonism \citep{Sidgwick}. Furthermore, \citet[pp. 16]{edgeworth1881} tried to reduce the distance between the two species of agents from two point of views, theoretical and empirical. From a theoretical point of view, Edgeworth claimed that the Pure Utilitarian ``might think it most beneficent to sink his benevolence towards competitors'', and that the pure egoistic might have need of a utilitarian calculus. This last point is of crucial importance for the building of a coherent theory in \textit{Mathematical Psychics}, as we will see below.

From a practical point of view, Edgeworth claimed that the two extremes of agents acting only on the basis of a utilitarian end or of pure self-interest are not a good approximation of ``concrete agents''. Instead,  
\begin{quote}
    ``the moral constitution of the concrete agent would be neither Pure Utilitarian, nor Pure Egoistic, but \foreignlanguage{greek}{μικτή τις} [of mixed kind]''. [...] For between the two extremes [...] there may be an indefinite number of impure methods; wherein the happines of others as compared by the agent (in a calm moment) with his own, neither counts for nothing, not yet `counts for one', but counts for a fraction.'' \citep[p. 16]{edgeworth1881}  
\end{quote}

Edgeworth, in fact, had a very clear understanding of the distinction between the theoretical simplification of a model -- ``the conception of Man as a pleasure machine'' \citep[p. 15]{edgeworth1881} -- and the complex psychological and moral motives of real people. This methodological awareness is crucial for correctly interpreting his ``first principle of Economics'', which he states as ``every agent is actuated only by self-interest'' \citep[p. 16]{edgeworth1881}.

This principle is the origin of the crass simplifications often attributed to Edgeworth's thought and usually adopted in handobooks of the history of economic thought.  However, if one considers the footnote attached to this first principle,\footnote{ ``\textit{Descriptions} rather but sufficient for the purpose of these tentative studies'' \citep[p. 16, note 2, italics in original]{edgeworth1881}.} it becomes clear that he viewed it not as a logical axiom, but as a simplified description of reality, useful for applying the calculus in a tractable way. He was isolating this trait to construct a formal model, viewing the self-interest axiom as a deliberate heuristic simplification for analytical clarity rather than a comprehensive description of reality. Ultimately, the profound challenge of reconciling this simplified model with the extreme complexity of real-world agents may well be the origin of the abrupt and, to many commentators, rather inexplicable shift in Edgeworth's focus in the years immediately following Mathematical Psychics from pure economic theory to the realms of probability and statistics \citep{baccini_2011}.

\textbf{Contract and competition.} The primitives of Edgeworth's Economical Calculus are then \emph{self-interested agents}, who may be individuals or
combinations of individuals; \emph{articles} of the contract may be
economic goods, but also property rights; \emph{actions} are what agents
do. All actions can be classified as \emph{war} or \emph{contract}
``according as the agent acts \emph{without,} or \emph{with,} the consent
of others affected by his actions''. The action of \emph{recontracting}
without the consent of others is war. ``The \emph{field of competition} with reference to a contract, or contracts, under consideration consists of all the individuals who are willing and able to recontract about the articles under consideration''. A ``\emph{perfect} field of competition'' is a field with perfect communication between individuals in which: (i) any individual is free to recontract with any out of an indefinite number of other agents; (ii) any individual is free to contract with an indefinite number, so each
article of contract is perfectly divisible and (iii) any individual is
free to recontract with another without the consent of any third party.

Edgeworth proposed two solution concepts \citep[pp. 16-19]{edgeworth1881}. The first one reads: ``A \emph{settlement} is a contract which cannot be varied with the consent of all the parties to it''. Or in modern jargon: a contract is a Pareto optimal allocation for the parties to the contract. The second one reads: ``A \emph{final settlement} is a settlement which cannot be varied by recontract within the field of competition'', i.e. with renegotiation with any or all of the parties outside the contract, but inside the field. This signifies, in modern jargon, that a final settlement lies in the \emph{core}. 
Therefore, and finally, ``a contract is \emph{indeterminate} when there are an indefinite number of \textit{final settlements}'' \citep[pp. 16-19, italics in original]{edgeworth1881}.

These definition are the basis for responding to the main research question of Edgeworth's Economical calculus : ``How far contract is indeterminate'' \citep[p. 20]{edgeworth1881}. This problem is central because indeterminacy acts as a widespread barrier to reaching contractual arrangements. The inquiry may therefore help to identify paths to overcome the critical impediments generated by it \citep[p. 20]{edgeworth1881}.

Edgeworth described the ``\textit{conclusions}'' he reached as follows:
\begin{quote}
``($\alpha$) Contract without competition is indeterminate, ($\beta$) Contract with \emph{perfect} competition is perfectly determinate, ($\gamma$) Contract with more or less perfect competition is less or more indeterminate.'' \citep[p. 20, italics in original]{edgeworth1881}
\end{quote} 

The solution for the so-called replica economy under perfect competition was largely celebrated as Edgeworth's most important contribution to economic theory (for all \citet{hildenbrand}). However, the focus on this solution and on the concept of equilibrium overshadowed his primary concern with the problem of indeterminacy which he considered the \textit{real} condition of economic relations, but also of ``the political struggle for power [and] the commercial struggle for wealth'' \citep[p. 16]{edgeworth1881}.

\textbf{Indeterminatess of contract.} According to Edgeworth, a contract is indeterminate when the number of contracting parties -- be they individual agents or \emph{combination} of agents such as trade unions in the labour market -- is limited. A low number of contracting parties leads to what he termed ``the absence of  competition'' which in turn causes ''the indeterminateness of agreements'' \citep[p. 449]{Edgeworth_1883_rev_jevons}. Edgeworth observed that this lack of competition was prevalent in contemporary society due to the rise of trade-unionism and cooperative association. From this, he inferred that the extent of indeterminateness would growth rapidly. Furthermore, he identified many fields characterized by the general absence of a mechanism like perfect competition and a prevalence of indeterminateness, noting its presence ``in international, in domestic politics; between nations, classes, sexes'' \citep[p. 50]{edgeworth1881}. 

Edgeworth concluded this line of reasoning by highlighting that his own theoretical findings threaten to undermine the very foundation of current economics: ``the reverence paid to competition'' \citep[p. 50]{edgeworth1881}. According to Edgeworth, economists have ``complacently acquiesced'' in the results of competition, viewing them as if they were the outcome of impersonal and impartial physical forces. There was no pretence that these results were morally just or humane, but they commanded a certain respect due to their ``majestic neutrality'' akin to the blind, unbiased forces of Nature. However, the ``field of competition'' often lacks both the ``continuity'' and the ``multiety of atoms'' (i.e., a large number of agents) that create the predictable uniformities we see in physics. When these conditions hold, the mechanism of competition lacks not only the ``regularity of law'' but also the ``impartiality of chance'', and it is no better than ``the throw of a die loaded with villainy''. Consequently, economics would truly be a ``dismal science'', and the reverence for competition would vanish \citep[p. 50, italics in original]{edgeworth1881}.\footnote{This is the complete quotation:\begin{quote}
    ``To impair, it may be conjectured, the reverence paid to competition; in whose results-as if worked out by a play of physical forces, impersonal, impartial-economists have complacently acquiesced. Of justice and humanity there was no pretence; but there seemed to command respect the majestic neutrality of Nature. But if it should appear that the field of competition is deficient in that continuity of fluid, that multiety of atoms which constitute a the foundations of the uniformities of Physics ; if competition is found wanting, not only the regularity of law, but even the impartiality of chance --the throw of a die loaded with villainy -- economics would be indeed a `dismal', and the reverence for competition would be no more.'' \citep[p. 50, italics in original]{edgeworth1881}
\end{quote}}

What solution, then, can be found to the problem of the indeterminateness of contract? In such a framework, a world characterized by indeterminacy risks devolving into a war of all against all, and this pervasive condition demands a universal solution. The way in which the problem is stated and solved by Edgeworth is rhetorically grandiose: 
\begin{quote}
``The whole creation groans and yearns, desiderating a principle of arbitration, an end of strifes.” \citep[p. 51]{edgeworth1881} 
\end{quote} 
His answer was that the solution is not to be found within the standard rules of economics -- in particular, competition -- but in a higher-order principle of arbitration inspired by utilitarianism.
In more prosaic terms and with specific reference to the labour market, \citet[449]{Edgeworth_1883_rev_jevons} two years later re-stated that same concept: when indeterminateness prevails, due to ``difference of interest between the two parties; there is required a principle of agreement''.

\textbf{The origin and meaning of arbitration.} 
The question of arbitration is a central node for understanding Edgeworth's intellectual path. Mirowski interpreted ``arbitration'' as an authoritative external intervention over the failure of the market: 
\begin{quote}
 ``outside arbitrators must \textit{impose} the utilitarian maximum. [...] The utilitarian maximum must then be justified on moral and non-economic grounds, and pleasure must be measurable, if the settlement is to be imposed by some external quasi-governmental body.'' \citet[p. 28]{mirowski1994}
\end{quote}
On a similar note, \citet{Creedy_1984} argues that the origin of the arbitration principle can be traced to Jevons's discussion on the indeterminacy of bargaining between isolated agents. Jevons indeed noted that ``indeterminate bargains [...] are best arranged by an arbitrator or third party'' \citep[p. 124]{Jevons_1871}. Creedy thus appears to read ``arbitration'' in \textit{Mathematical Psychics} in the same way as in Jevons, that is, as a call for an external authority to intervene between parties struggling to reach an agreement.

Creedy's interpretation, however, is at odds with Edgeworth's critical appraisal of Jevons's idea of arbitration, which he reiterated in two reviews of Jevons published in \textit{The Academy}.\footnote{The existence of these two reviews was not known until \citet{baccini2003}.} In the first one, \citet{edg_rev_jevons} acknowledged that Jevons correctly identified the problem of indeterminacy in contracts between trade unions and employers and also recognized that ``the smoothing and rounding agency of conciliation'' can help reach an agreement. However, he reproached Jevons for not having ventured to define a ``principle of arbitration''" One year later, Edgeworth returned to this point, again criticizing Jevons for failing to establish a principle of arbitration in defining industrial partnership agreements \citep{edg_rev_jev_methods}.

In fact, Edgeworth used ``arbitration'' to indicate the ``basis'' or ``principle'' that agents or coalition of agents adopt to negotiate an agreement and concile their contrasting interests. This use of the term was largely diffused when he wrote \textit{Mathematical Psychics}, in particular in debates surrounding the relationships between employers and their associations, and workers and trade unions -- an issue that would later be conceptualized as ``collective bargaining'' \citep{Webb}. 

In 1881, the main available contribution to this issue was made by \citet{Crompton}. \citet[p. 134]{edgeworth1881} cited Crompton in reference to the ``\textit{combination} of tenants against landlords, which the present crisis in Ireland is thought to involve''.  Edgeworth considered Crompton as the authority indicating ``the need of some principle of arbitration''. Edgeworth's reference points to \citet[pp. 81-82]{Crompton} who argued that for an arbitration to be satisfactory and not a mere compromise, it must be guided by a fixed, albeit somewhat arbitrary, rule. Crompton's proposed method was to establish a practical ideal based on a historical period where wages and profits were deemed satisfactory, and then use a fixed ratio between prices and wages from that period as a benchmark for future settlements. He concluded that this was the only ``rational theory'' for such arbitrations.\footnote{The full quotation is 
\begin{quote}
``The truth that these arbitrations will not be satisfactory unless they are based on all the facts, relative to prices, profits, and the demand and supply of labour, and are determined by some arbitrary but fixed rule, which alone can prevent the arbitration from being a mere compromise. Generally this is a ratio agreed upon between prices and wages. A certain date is fixed upon, at which time the wages were considered to be satisfactory to the men and the profits to the master. This becomes the practical ideal, a just and desirable condition and the arbitrator's aim, when acquainted with all the facts, is to approximate to this ratio, as far as he can do so without injury or injustice. I do not believe that there is any other rational theory of these arbitration'' \citep[pp. 81-82]{Crompton}.    
\end{quote}}
Crompton did not use ``arbitration`` to indicate the intervention of an external authority, but rather the set of principles guiding a settlement. This usage was later characterized as paradigmatic of an ``old'' meaning by Beatrice and Sydney Webb:  
\begin{quote}
    ``until quite recently, no clear distinction drawn between Collective Bargaining, Conciliation, and Arbitration. Much of what is called Arbitration or Conciliation in the earlier writings on the subject amounts to nothing more than organised Collective Bargaining.''\citep[p. 222]{Webb}
\end{quote}
\citet{Webb} reinforced the evidence of this old use of the term by citing a talk by Antony John Mundella, a key figure in Victorian social reform. As a  member of the parliament, minister, and pioneering advocate for compulsory education and child labour laws, Mundella was instrumental in establishing a groundbreaking system of voluntary arbitration and conciliation to resolve industrial disputes. According to Mundella:
\begin{quote}
``The sense in which we use the word [arbitration] is that of an arrangement for open and friendly bargaining in which which masters and men meet together and talk over their common affairs openly and freely.''  \citep[quoted in][p. 223]{Webb}
\end{quote}
Hence, Edgeworth's ``arbitration'' was neither a generic call for an external authoritarian intervention between combinations of economic actors such as employers and employees, nor a call for an arbiter to arrange otherwise indeterminate contracts. It referred, rather, in line with the contemporary  use of the term, to the very definition of the basis or, better, the principle upon which parties could reconcile their contrasting interests, negotiate an agreement, and write a contract. 

\textbf{The principle of arbitration.}  The question at this point is ``where a world weary of strifes seek a principle of arbitration?'' \citep[p. 51]{edgeworth1881}. Edgeworth discussed the possible replies to this question in two dense and overlooked pages \citep[pp. 51-53]{edgeworth1881} where he fundamentally rejected abstract concepts of justice and fairness as viable principles for arbitration. He contended that while these ideas are morally elevating, they fail to provide a definite, quantitative criterion for action. They are ``charming'' but offer no workable method to answer practical questions, such as how to precisely divide a joint product between collaborators. For instance, there is no reason why one cooperative distribution model is inherently more ``just'' than any number of other possible arrangements.\footnote{Edgeworth referred here to George Jacob Holyoake (1817-1906), a British secularist and co-operator, to whom he attributed a specific model for distributing the product within a cooperative. In his model, the operative workers would receive the net product of their enterprise. From this total sum, a salary would then be paid to the manager. Edgeworth's criticism was not of the model's specifics, but of its foundational logic. He argued that Holyoake's principle is no more inherently ``equitable'' than an indefinite number of other conceivable distribution schemes (e.g., the operatives taking an arbitrary agreed fraction, with the manager taking the remainder).} Without a concrete standard, justice and fairness are merely vague and non-committal utterances.

This leaded Edgeworth to argue that justice requires a more precise guiding light: utilitarianism. However, Edgeworth was skeptical that this moral principle alone can overcome the ``controlless core of human selfishness'' in domains like war and trade. His analysis, echoing  Sidgwick, acknowledges two supreme and irreconcilable conflicting principles: Egoism and Utilitarianism. Edgeworth admitted that religion might reconcile them, but he set this option aside for his inquiry. Focusing on the lower elements of human nature, his goal was instead to find a ``more earthy passage'' leading  from self-interest to the ``principle or at least, the practice, of utilitarianism''.

It is precisely at this juncture in Edgeworth's reasoning that ``sympathy'' emerges as the solution, that is both analytical and conceptual. Analytically, Edgeworth modeled sympathy by introducing in an agent's utility function the utility of the other agent, weighted by a coefficient of ``effective sympathy''. He then argued that the original contract curve restricts ``between narrower limits'' and, as the coefficients of sympaty increase, the ``\textit{the contract- curve narrows down to the utilitarian point}'', that is to the point where the sum of the utilities of the two agent reaches a maximum.\footnote{\citet{Collard} reconstructs Edgeworth's reasoning using modern notation and assesses its validity.}

Conceptually, the emergence of the utilitarian point as the contractual solution between altruistic pleasure machines permits to satisfy at the same time ``the sympaty (such as it is) of each with all, the sense of justice and utilitarian equity'' \citep[p. 51]{edgeworth1881}. It is a rather puzzling phrase. The second part refers clearly to what Edgeworth previously noted: that the sense of justice is only justified if it is enlightened by the utilitarian principle. The first part refers instead to the idea that when an agent has even a minimal inclination\footnote{In the corresponding footnote, \citet[p. 54 n. 1]{edgeworth1881} implicitly invokes Lucretius's concept of the \textit{clinamen} -- a slight swerve -- to indicate a deviation from the ``rectilinearity of economic man''. Lucretius used the word to indicate ``a minimal deviation by atoms from their straight course that occurs randomly, at no specific time or place''. For Lucretius, this random, uncaused deviation was essential to break causal determinism and explain free will, as voluntary action would otherwise be trapped within an infinite causal chain \citep{sep-lucretius}.} towards others, this is sufficient, when formalized in the utility function, to narrow the contract curve toward the utilitarian solution, called now utilitarian equity. As we have already seen in NOME, Edgeworth did not assume perfect altruism for real-world agents precisely because he viewed humanity's evolution as incomplete; a perfectly altruistic utilitarian agent is only a potential endpoint, whereas actual agents exhibit only the nascent, imperfect sympathy of a species still progressing toward that ideal.

For illustrating the path conducing from indeterminateness to the final utilitarian contractual solution between moderately altruistic economic man, Edgeworth discussed the case of the bargaining between an entrepreneur and a worker. They have different and contrasting aims. He started by assuming that they arbitrate by adopting ``some principle of \textit{doctrinaire justice} -- some metaphysical dogma, for instance of equality'' \citep[p. 54, italics in original]{edgeworth1881}.
Edgeworth argued that there is no reason to assume a pre-defined ``fair'' division between a worker and an entrepreneur is either utilitarian or efficient. Their self-interest will inevitably push them from any such fair arbitrary point to a settlement on the contract-curve.

Edgeworth's reasoning continues with a complex passage. He supposed that after repeated negotiations a range of possible efficient outcomes -- all lying on the contract curve -- would emerge, leaving the parties facing an indeterminacy problem. These efficient outcomes are in a reverse order of desirability for each party and appear equiprobable.\footnote{\citet[p. 82]{creedy1986} interprets this specific passage by claiming that Edgeworth viewed distributive justice in terms of choice under uncertainty. However, Creedy collapsed the problem of distributive justice, that is the object of Edgeworth's utilitarian calculus, with the defintion of the principle of arbitration determining the contract when competition is lacking. Edgeworth was illustrating a simple bargaining euristic, not the general foundation of distributive justice. Moreover, Creedy overextends this reading by collapsing the notion of equiprobability -- only hinted at here -- with the strong and formal use of it found in Edgeworth's subsequent contributions.} Edgeworth claimed that ``rather than resort to some process which may virtually amount to tossing up, both parties may agree to commute their chance of any of the arrangements for the certainty of one of them'' \citep[p. 55, italics in original]{edgeworth1881}. In fact, Edgeworth does not discuss here distributive justice, that is the object of utilitarian calculus, but the application of a principle of arbitration. 

Edgeworth considered two possible scenarios: one in which the parties choose the utilitarian arrangement directly, and another in which the parties agree to ``split the difference'' and settle on a central point, the ``quantitative mean'' among the options. He then made a critical connection: this very notion of a mean is the manifestation of a rudimentary, ``implicit'' sense of justice, poised to evolve into the sophisticated, ethical ideal of the ``qualitative mean'' of ``utilitarian equity''.\footnote{\citet{edgeworth_1889, edgeworth_1889_1} reiterated this reasoning in reference to recent literature.}

This process occurs because, as Edgeworth explained, \begin{quote} ``in the neighbourhood of the contract-curve \textit{the forces of self-interest being neutralised}, the tender power of sympathy and right would become appreciable.''\footnote{Edgeworth illustrated this with an analogy drawn from physics: in the neighborhood of contract curve, sympathy become visible because the force of self-interested are neutralized, ''as the gentler forces of the magnetic field are made manifest when terrestrial magnetism, by being opposed to itself, is eliminated''\citep[p. 56]{edgeworth1881}. }\citep[p. 56] {edgeworth1881}\end{quote} 
Thus, a simple bargaining heuristic becomes the embryo from which full-fledged utilitarian justice can bloom, but only after the force of self-interest has created a stalemate where finer moral considerations can finally be felt.

\textbf{The principle of arbitration is utilitarianism.} Only after embarking on this lengthy process of justification did Edgeworth finally define the principle clearly:
\begin{quote}
    ``\textit{competition requires to be supplemented by arbitration, and the basis of arbitration between self-interested contractors is the greatest possible sum-total utility.}'' \citep[p. 56, italics in original]{edgeworth1881}
\end{quote}
In summary, with his “economic calculus”, Edgeworth identified two complementary processes through which economic men reach agreements: perfect competition and settlement according to the utilitarian principle of arbitration. 
In the first case, the one prevalently celebrated by subsequent literature, the competition determines the price and the equilibrium in the market. 
In the second case, the utilitarian solution of maximizing social welfare in a bargaining context is not the outcome of an external authority's intervention, but rather emerges as contractual solution between self-interested agents, be they individuals or coalitions. This, however, raises a problem at the core of Edgeworth's theory: while the self-interested pleasure aims to maximize their own utility, accepting the utilitarian solution as an agreement may conflict with this very objective. Indeed, even if a solution that maximizes total utility is theoretically efficient, it will not be voluntarily agreed upon by individuals if it leaves any of them worse off than they were before any exchange took place. In modern jargon, even if the utilitarian solution lies on the contract curve, it cannot be realized by self-interested agents unless it constitutes a Pareto improvement over their pre-contractual position.\footnote{This tension is likely the reason why the prevalent interpretation of Edgeworth's ``arbitration'' considers external intervention as necessary, and dictated by reasons that are not based on economics.}
As we have seen, the emergence of the utilitarian solution is due to the altruistic nature of agents, which in turn is due to the evolution of human beings, as Edgeworth had extensively discussed in NOME. 

\textbf{The utilitarian calculus}. It is at this juncture that the ``economical calculus'' gives way to the ``utilitarian calculus''. The former has shown that (i) under perfect competition -- a very specific case -- a determinate equilibrium emerges by modeling individuals as mere self-interested pleasure machines; and (ii) in the absence of competition -- the real and widespread case --this representation must be abandoned. To solve the problem of indeterminacy, we must instead model agents as being only partially self-interested and partially altruistic, at least at the present stage of evolution. It is these more realistic agents who seek an agreement guided by the utilitarian principle of arbitration.

This realization allowed Edgeworth to return to his grand vision of the hedonical calculus, which he re-proposed it in its entirety under the new name of the ``utilitarian calculus''. This framework serves as a generic description of the distributive process that leads to the maximization of social welfare. 

As Edgeworth himself he would note years later, this is the precise point where economics becomes ``blurred and obliterated by the action of political and ethical elements``.\footnote{The full quotation is: \begin{quote}
``But when we are not entitled to assume that each individual is independently trying to make the best bargain for himself -- when combinations act in concert -- the exact character of economic science becomes blurred and obliterated by the action of political and ethical elements. What is now required is the sort of considerations which regulate a commercial treaty between two communities, rather than the laws appropriate to a perfect market''. \citep[402]{Edgeworth_rev_price}
\end{quote}}
Consequently, the study of distribution and welfare maximization transcends the traditional boundaries of economics, concerning itself not merely with economic actors, but with all sentients. This expanded analytical task, which he had previously defined in its entirety, was accomplished by modeling sentients through his ``quasi-Fechnerian'' formula. 

\section{Conclusion}
This extended journey through Edgeworth's texts on distribution and economic choice has revealed a grand intellectual design that transcends the narrow focus often attributed to him by economists. His overarching goal was to unificate moral sciences thorugh the application of mathematics. 

The theoretical project outlined in this initial phase of Edgeworth's work aimed to reconfigure utilitarianism as a precise scientific discipline based on mathematics, psychophysics, and evolutionary biology. Its model of sentient beings as quasi-Fechnerian machines integrates political theory with natural law, arguing that anti-egalitarianism is not a normative principle but an empirical outcome of maximizing welfare among unequally endowed beings. This abstract finding is powerfully instantiated in the figure of Mr. Pongo. His distinct position on the evolutionary scale exemplifies the theory's core tenet: the unequal weights assigned to sentient individuals in the social welfare function are not derived from an external theory of justice but are dictated by measurable, natural, biological disparities in the capacity for happiness.

What holds true for sentient beings across different evolutionary scales continues to apply even when focusing solely on human societies. The different capacities of human individuals are the result of the evolutionary process, and the rule of welfare maximization once again points toward an unequal distribution, as a result of these natural differences among races, sexes, and social classes. The system, in essence, uses the apparent neutrality of science to naturalize, justify, and reinforce pre-existing social hierarchies.

This project led him eventually to devise novel analytical tools for economics, which are rightly celebrated in the subsequent literature. However, one aspect that has been insufficiently emphasized is Edgeworth’s clear understanding of the need to construct different models for different problems. The conventional view of him as a mere father of \textit{homo economicus} reduces his work to a neoclassical caricature, even a crude one. In reality, as we have seen, Edgeworth employed at least three distinct models to represent the individuals in his analyses.

For his hedonical (later utilitarian) calculus, Edgeworth executed a veritable model transfer \citep{Herfeld}. He imported the representation of sentient beings in terms of stimulus/response from the psychophysics of his time into the ethical calculus -- note carefully, not in economics -- by representing them as ``quasi-Fechnerian'' sentients. He was quick to point out that this representation was consistent with an evolutionary view of these very beings. 

For economical calculus, Edgeworth required a less grand representation. For the particular, and not so useful, case of perfect competition, he contented himself with the model of agents as purely self-interested pleasure machines. As soon as the analysis becomes more complex -- namely, when moving from the abstraction of perfect competition to the reality of bargaining -- the appropriate model shifts to that of an agent who is only moderately self-interested and only moderately altruistic. Again, this representation was framed as being consistent with the current stage of human evolution.

In all three cases, Edgeworth was acutely aware that these were simplifying model constructs, recognizing that the complexity of sentient and human action far exceeds what any mathematical model can capture. His justification for their use stemmed from his beloved ``Law of Error'': these schematic representations could be conceived as capturing the average of a vast underlying complexity -- the ``multiety'' of real  ``atoms''. As he would later elaborate in an article dedicated to the rationale of exchange \citep{edgeworth1884a} --an explanation that commentators have dismissed as a ``rather forced analogy'' \citep[p. 96]{newman1987} -- this was a deliberate methodological choice to render the immense variety of human nature tractable for scientific inquiry. For Edgeworth, the use of deterministic mathematical models is subsumed under the principle that they are designed to capture the average tendencies of a reality governed by statistical regularities \citep{baccini_2011}.

Economists who discussed Edgeworth's work are often unsettled by the clarity of his utilitarianism and the harsh, anti-egalitarian, and eugenic conclusions he derives from it.\footnote{In direct opposition to this reductive view of Edgeworth's utilitarianism stands the interpretation that it forms the basis of modern contractarian utilitarianism. Notably, \citet{Rawls} himself refers to Edgeworth's contractual solution, while \citet{Creedy_1984} considers him a precursor to the results of Harsanyi.} The discomfort was evident from Marshall's earliest review to modern scholars \citep[e.g. among the others][]{Hicks, creedy1986, newman1987, screpanti_zamagni}. There is likely a sense of embarrassment in discovering that many foundational tools of microeconomics emerged from an intellectual milieu that appears unpalatable, at least when explicitly acknowledged. No less discomfiting, perhaps, should be the realization that contemporary ``happiness economics'' itself has roots in this same soil \citep{eceiza}.

In this regard, Schumpeter is the most explicit critic. He wrote:
\begin{quote}
``First, I mention [Edgeworth's] utilitarianism, which strongly asserted itself from the beginning and looked so incongruous in a man whose mind was nothing if not `cultured'; it did much to keep alive -- quite unnecessarily -- the unholy alliance between economics and Benthamite philosophy [...]. But let me also repeat that in his case, as in that of Jevons, we can leave out the utilitarianism from any of his economic writings without affecting their scientific contents.'' \citep{Schumpeter}
\end{quote}

However, this very operation -- the surgical removal of utilitarianism from Edgeworth's thought -- leaves standing only the tools of the ``economical calculus''. According to their own author, these tools are certainly not sufficient for solving the problems of economic welfare, let alone for explaining reality. Indeed, this surgical removal of utilitarianism kills the patient. Or, more precisely, it reduces his grand intellectual project to a skeletal framework that contemporary economist can safely acknowledge as a precursor, and a Whiggish historiography of economics can then comfortably celebrate.

A final consideration concerns the trajectory of Edgeworth’s work. With rare exceptions -- such as the works of \citet{mirowski1994} and \citet{baccini_2011} -- the historiography has struggled to understand why he appeared to abandon economic problems in the years following the \textit{Mathematical Psychics}. Alternative explanations for this shift cite the work's lack of success, its mixed contemporary reviews, and Edgeworth's own search for an intellectual hero \citep{stigler1978, newman1987}. In reality, all these perspectives misconstrue his intellectual project. In 1881, economics occupied only a few pages in a much broader and well-defined program: the application of mathematics to the moral sciences. The contradiction between the regularity of aggregate results and the unpredictability -- the ``multiety of atoms'' -- of individual behaviour became the primary object of his subsequent research in probability and statistics. It was only after achieving academic success with his appointment to the Drummond Chair of Political Economy at Oxford in 1890 that Edgeworth would significantly return to the field.

\theendnotes 

\bibliographystyle{apalike}

\bibliography{Bibliography.bib} 

\end{document}